**Apparent energy-speed relationship poses no challenge to Bohmian mechanics**

*Matthew Dickau,* matt.dickau@gmail.com

*August 2025 (v1)*

A recent article [1] claims to measure the speed of quantum particles in the classically forbidden regime where the energy of the particles is lower than the local potential, and further claims that the results of this experiment challenge Bohmian mechanics. But this interpretation of the experiment is incorrect (and dubious even in the context of ordinary quantum mechanics). A proper analysis of the system from a pilot-wave perspective shows that it predicts the same distribution of particle positions (and so the same experimental results) as ordinary quantum theory. The "speed" measured by the experiment in this regime is fictitious.

**Explanation of the Experiment**

The experiment [1] of Sharoglazova et. al. cleverly uses photons in an optical cavity (effectively, light reflecting between two mirrors) to physically simulate non-relativistic quantum particles. Projecting out the motion normal to the mirrors, the residual planar motion of the photon has the same behaviour as a massive quantum particle in 2 dimensions, and by etching nano-scale structure onto one of the mirrors (varying the local distance between the mirrors), the experimenters can control the effective potential in which the simulated particle moves (with the decrease in distance proportional to the increase in potential energy). Moreover, since the mirrors allow some light transmission, the photons eventually escape the cavity, enabling a measurement of their position.

In this case the optical cavity takes the form of a 1-dimensional waveguide (a groove) in which the photons can travel until they reach a step up in the potential (the groove gets shallower). At this point a second waveguide is introduced parallel to the first, and the photons which are transmitted through the step can tunnel back and forth between the first (main) and second (auxiliary) waveguides. The experimenters consider the rate at which the particles transition between the grooves as a clock, enabling them to infer the speed of the particles from their spatial distribution along the two grooves (or so it seems).

For a highly simplified model of this, consider the case where the particle can only have two states, being in the main waveguide or the auxiliary waveguide. Then the wavefunction of the particle is just two numbers, $\psi_m$ and $\psi_a$, governed by the equations (here we use units where $\hbar = 1$):

$$i\frac{d\psi_m}{dt} = J_0\psi_a\,, \qquad i\frac{d\psi_a}{dt} = J_0\psi_m \tag{1}$$

If the initial condition is $\psi_m = 1$ and $\psi_a = 0$, the solution is:

$$\psi_m(t) = \cos(J_0 t) \ , \qquad \psi_a(t) = -i\sin(J_0 t) \tag{2}$$

So that the probability for the particle to be found in the auxiliary waveguide is $p_a = |\psi_a|^2 = \sin^2(J_0 t)$, which is approximately $(J_0 t)^2$ for early times. If the particle is travelling along the waveguide, this can be used to find the speed by writing $t = x/v$, so that $v$ can be inferred:

$$p_a = \left(\frac{J_0 x}{v}\right)^2 \tag{3}$$

Introducing the 1-dimensional spatial degree of freedom, the model can be expanded so that the particle wavefunction consists of two functions of position along the waveguides, one representing the main waveguide, and the other representing the auxiliary. Now the Schrodinger equation for the system is:

$$\begin{aligned} i\frac{\partial \psi_m}{\partial t} &= -\frac{1}{2m}\frac{\partial^2 \psi_m}{\partial x^2} + V_m\psi_m + V_i\psi_a \\ i\frac{\partial \psi_a}{\partial t} &= -\frac{1}{2m}\frac{\partial^2 \psi_a}{\partial x^2} + V_a\psi_a + V_i\psi_m \end{aligned} \tag{4}$$

Where the potential functions have the form:

$$\begin{aligned} V_m(x) &= \begin{cases} 0, & x < 0 \\ V_0 - J_0, & x \geq 0 \end{cases} \\ V_a(x) &= \begin{cases} \infty, & x < 0 \\ V_0 - J_0, & x \geq 0 \end{cases} \\ V_i(x) &= \begin{cases} 0, & x < 0 \\ J_0, & x \geq 0 \end{cases} \end{aligned} \tag{5}$$

The infinite potential for the auxiliary waveguide on $x < 0$ implies $\psi_a(x,t) = 0$ for that region. Of course, this is an idealization of the true experimental setup, simplifying details of the potential landscape and neglecting the processes by which photons are introduced into and escape from the optical cavity (effectively, decaying out of the system).

Using millimeters and nanoseconds as our basic length and time units and setting $\hbar = 1$, the system parameters were $V_0 \approx 817\ \mathrm{ns}^{-1}$, $J_0 \approx 40\ \mathrm{ns}^{-1}$, and $m \approx 65.9\ \mathrm{ns}\cdot\mathrm{mm}^{-2}$. The decay rate is $\gamma \approx 3.7\ \mathrm{ns}^{-1}$, corresponding to a particle lifetime of $1/\gamma \approx 0.27$ ns. This is the same order of magnitude as the time expected for the particles to cross the experimental setup. Later, we will consider the effect of the finite particle lifetime on the system's dynamics, and its impact on the discussion.

A laser pulse introducing photons into the system is located at a potential ramp at the beginning of the main waveguide, allowing control over the initial energy of the particles. In the experiment the pulse was 26 ns in duration, long enough for the particle distribution to build up into a quasi-stationary state. For a theoretical approximation, we assume a stationary wavefunction (again, neglecting the finite particle lifetime):

$$E\psi_m = -\frac{1}{2m}\frac{\partial^2\psi_m}{\partial x^2} + V_m\psi_m + V_i\psi_a$$
$$E\psi_a = -\frac{1}{2m}\frac{\partial^2\psi_a}{\partial x^2} + V_a\psi_a + V_i\psi_m \tag{6}$$

Solving these equations (see Appendix A), we get different behaviour depending on whether the quantity $\Delta = E - V_0 + J_0$ is positive or negative. It is assumed in [1] that $|\Delta/J_0| \gg 1$, and the regime where $J_0$ and $\Delta$ are similar in magnitude is only briefly mentioned. See [6][7] as well as the appendix here for a more precise analysis.

If $\Delta > 0$, the particle classically has enough energy to be transmitted into the region of increased potential, and the wavefunction continues to propagate there. If $\Delta < 0$, this is the classically forbidden regime: the particle tunnels into the step-up region even though this makes its kinetic energy locally negative, and its wavefunction decays exponentially as it gets further in.

The proportion of particles found in the auxiliary waveguide is:

$$p_a(x) = \frac{|\psi_a(x)|^2}{|\psi_m(x)|^2 + |\psi_a(x)|^2} \tag{7}$$

In the solutions, this proportion is related to a quantity defined by:

$$k_1 = \frac{J_0}{\sqrt{2|\Delta|/m}} \tag{8}$$

In both cases $\Delta > 0$ and $\Delta < 0$, for small $x$, $p_a(x) \approx (k_1 x)^2$ and we extract the velocity, up to a sign, from equation (3):

$$v \approx \sqrt{2|\Delta|/m} \tag{9}$$

The experimental results bear out the relationship $p_a(x) \approx (k_1 x)^2$ for small $x$, with $k_1$ defined as above. The authors of [1] take their results to be indicative of a successful measurement of the speed $v$. (As noted in [6], if we do not assume that $|\Delta/J_0| \gg 1$, the relationship between $p_a$ and $|\Delta|$ deviates slightly from what is laid out above and in [1], in a way that better agrees with the experimental data, but this has little bearing on the interpretation of the experiment.)

**Interpretation of the Experiment**

Even before considering how this relates to Bohmian mechanics, there are reasons to view this measurement of the "speed" of the particle with some suspicion. In orthodox quantum mechanics, particles do not have well-defined positions, and therefore have neither trajectories through space-time nor velocities (the slopes of those trajectories) nor speeds (the magnitudes of those slopes) in the usual sense. An approximate velocity may be defined if the particle wavefunction is described by a localized wave packet: in that case, the particle trajectory is "blurry" but can still be made sense of over large enough time and length scales.

To see what the approximate trajectory looks like in the context of this experiment, we must re-conceptualize it to think of the evolution of an incident wave packet, rather than a stationary state. [2][3] The process begins with a wave packet (very wide relative to its wavelength) travelling towards the step potential in the main waveguide. What happens next depends on the sign of $\Delta$. (In the following we assume $|\Delta/J_0| \gg 1$ for simplicity.)

If $\Delta > 0$, when the leading edge of the wave packet reaches the step potential, part of it is reflected, with a lower amplitude than the incident packet. This creates a region before the step where the incident and reflected wave packets overlap and interfere with each other. Meanwhile, another part of the incident wave packet is transmitted into the higher potential region. Eventually the trailing edge of the incident packet reaches the step-up, and the reflected and transmitted wave packets are clearly separated, travelling in opposite directions. The reflected wave packet has the same speed as the incident packet, while the transmitted packet is travelling more slowly. Even during the reflection, the edges of the wave packets have relatively well-defined velocities, since (because the packets are very wide relative to their wavelengths) their shapes do not appreciably distort. See Figure 1 for an image from a simulation of this process.

Things are different if $\Delta < 0$. In that case, the incident packet reflects off the step-up, but (once the leading edge has passed) the reflected packet has the same amplitude compared to the incident packet. The leading edge of the incident packet penetrates the higher potential region, but exponentially decays over the penetration distance. Eventually the trailing edge of the incident packet reaches the potential step, and the portion of the wavefunction in the classically forbidden region is rejected, rejoining the trailing edge of the reflected packet as it travels away. Here there is no transmitted wave packet to which we can assign a velocity, and even during the (extremely brief) periods that the wavefunction is penetrating or being rejected from the high potential region, its edge does not have a well-defined speed due to its shape being distorted by the exponential decay. See Figure 2 for an image from a simulation of this process.

The probability current gives another indication of the difference in behaviour between the classically allowed and classically forbidden regimes. When $\Delta > 0$, there is a net probability current in the transmitted wave packet. But for $\Delta < 0$, there is no net probability current in the high potential region, except during the transient phases at the beginning and end of the reflection process. Sharoglazova et. al. choose to interpret this zero value as a balance between probability fluxes entering and leaving the high potential region, but this is an assumption without a solid physical warrant. (Note that the net probability current is also zero in the region of overlap between the incident and reflected wavepackets, in the $\Delta < 0$ case.)

In both cases, some of the particle wavefunction in the high potential region tunnels from the main waveguide to the auxiliary waveguide, developing the spatial structure of $p_a(x)$ from which the experimenters infer their speed measurement. But it is only in the $\Delta > 0$ case that there is any recognizable motion in the wavefunction to which we may associate this speed.

Sharoglazova et. al. defend their interpretation of the measured quantity $v$ as an actual speed of the particles in the following way:

"[T]his quantity [$v$] determines the spatial length scale of the population build-up in the coupled waveguide system – an effect that genuinely represents a spatio-temporal phenomenon, distributing particles in space ($x$) relative to a temporal reference ($J_0$). We regard this as a property that can be attributed only to a form of motion." [1]

The use of $J_0$ as a temporal reference is justified from the behaviour of the simple two-state system considered above. However, when we expand to the 1-dimensional model, the main/auxiliary degree of freedom becomes entangled with the position degree of freedom, due to the potential having a different shape in the two waveguides. As a result, it is not guaranteed that the main/auxiliary degree of freedom exhibits the same temporal behaviour as it would if it were isolated. Indeed, since the quantum state in the experiment is (quasi-)stationary, while the oscillatory state (2) for the two-state system is not, it manifestly does not have the same temporal behaviour. Moreover, while for $\Delta > 0$ the main/auxiliary degree of freedom shows oscillatory variation with position, for $\Delta < 0$ it is not oscillatory, undermining the justification for considering the population hopping process as a reliable clock in this regime.

**The Experiment and Bohmian Mechanics**

In Bohmian mechanics, the wavefunction evolving according to the Schrodinger equation is associated with particles travelling along definite trajectories. The Bohmian particles describe the microscopic structure of the observable macroscopic world. Because of this, there is no need (at least at the fundamental level) for measurement postulates or the Born rule: these can be derived from the motions of the particles under the guidance of the wavefunction. There is no wavefunction collapse: the wavefunction always obeys Schrodinger's equation. And there is no measurement problem: measurements have definite outcomes because the particles always have definite positions.

Just as the Hamiltonian operator can have different forms depending on the system of interest, the guidance law that determines the motions of the particles can have different forms. In the case of this experiment, the configuration space of the system has two discrete sectors (the main and auxiliary waveguides). Here the guidance law must include both continuous deterministic motion within each sector, as well as discrete stochastic jumps between the sectors. This type of motion for a Bohmian system was first considered (to my knowledge) for modelling particle creation and annihilation [4], where the different sectors of configuration space correspond to different particle numbers. But the same techniques may be used for any configuration space with discrete sectors.

Note that for this case, the stochastic process is present because of how we are modelling the system. If we were to model the optical cavity as a 2-dimensional potential landscape rather than as two coupled 1-dimensional landscapes, the Bohmian guidance law would revert to its more familiar form, including only continuous deterministic motion. Effectively, the stochastic process arises from neglecting the unknown position of the particle along the width of the waveguide, which is what determines when it tunnels through the barrier between the two grooves in the 2-dimensional model. (However, stochastic processes may be fundamental features of other models, not arising from simplification: particle creation and annihilation may be such a case.)

Thus, for this system, the particles have Bohmian velocities:

$$v_m = \frac{1}{2mi|\psi_m|^2}\left(\psi_m^* \frac{\partial \psi_m}{\partial x} - \frac{\partial \psi_m^*}{\partial x}\psi_m\right), \qquad v_a = \frac{1}{2mi|\psi_a|^2}\left(\psi_a^* \frac{\partial \psi_a}{\partial x} - \frac{\partial \psi_a^*}{\partial x}\psi_a\right) \tag{10}$$

As well as Bohmian jump rates:

$$\sigma_m = \frac{[iV_i(\psi_m^*\psi_a - \psi_a^*\psi_m)]^+}{|\psi_m|^2}, \qquad \sigma_a = \frac{[iV_i(\psi_a^*\psi_m - \psi_m^*\psi_a)]^+}{|\psi_a|^2} \tag{11}$$

Where $[f]^+ = f$ if $f > 0$, otherwise $[f]^+ = 0$. What these quantities mean is that if the particle is in the main waveguide at location $x$, it has probability $\sigma_m(x)dt$ of jumping to the auxiliary waveguide (at the same location $x$) in an infinitesimal time interval $dt$, and if it does not jump, it proceeds a distance $v_m(x)dt$. (Similarly, if it is in the auxiliary waveguide, using $\sigma_a(x)dt$ and $v_a(x)dt$.)

Under these motions, any ensemble of particles for this system, described by probability distributions $\rho_m(x)$ and $\rho_a(x)$, will evolve according to:

$$\begin{aligned} \frac{\partial \rho_m}{\partial t} + \frac{\partial}{\partial x}(\rho_m v_m) = \sigma_a \rho_a - \sigma_m \rho_m \\ \frac{\partial \rho_a}{\partial t} + \frac{\partial}{\partial x}(\rho_a v_a) = \sigma_m \rho_m - \sigma_a \rho_a \end{aligned} \tag{12}$$

Here, the left-hand sides represent the flow of probability within each waveguide, and the right-hand sides represent the sources and sinks caused by probability transitioning between the two waveguides.

Now, it can be shown from the Schrodinger equation for this system that $(\rho_m, \rho_a) = (|\psi_m|^2, |\psi_a|^2)$ is a solution to the probability evolution equations (12) (see Appendix B). And this means that if an ensemble of particles, all with the same wavefunction, start off distributed according to $|\psi|^2$, then they will remain $|\psi|^2$-distributed for all times (even as both the wavefunction and the particle positions evolve).

There are arguments for why, in Bohmian mechanics, we should expect the particles to be $|\psi|^2$-distributed initially. [5] Here we may simply take it as an additional postulate (known as the quantum equilibrium hypothesis). The point is that with this postulate in place for the initial condition, the particles always remain $|\psi|^2$-distributed, and therefore predictions for the particle positions in Bohmian mechanics are always, necessarily, the same as the predictions for particle positions (and so for the results of this experiment) in ordinary quantum mechanics.

All of this is to say that this experiment does not falsify Bohmian mechanics. Bohmian mechanics predicts the same observed structure for $p_a(x)$ as ordinary quantum theory.

It is interesting to note that this experimental method accurately measures the Bohmian velocity in the classically allowed, $\Delta > 0$ regime (unless there is also some leftward propagating component in the high-potential region, such as a reflection from the end of the coupled waveguide system; in that case they would differ). But in the forbidden $\Delta < 0$ regime – if we ignore particle losses – both the Bohmian velocities and jump rates drop to zero, once the transient phase of the reflection process has passed and the quantum state becomes stationary. (In the next section we will consider the effect of particle losses.)

But again, the (perhaps unintuitive) “frozen” behaviour of the Bohmian particles is not refuted by this experiment. Once the spatial structure is built up in the transient phase, motion is unnecessary for the particles to remain $|\psi|^2$-distributed (and so reproduce the experimental results), since $|\psi|^2$ isn’t changing.

**Dwell Time and Particle Losses**

The finite lifetime of the particles adds a complication to the above discussion, which we will now consider along with a related objection from [1]. However, it does not change the overall conclusion, but rather further demonstrates the coherence of Bohmian mechanics.

Sharoglazova et. al. claim that the "dwell time" (the average time that particles supposedly spend trapped in the classically forbidden region before being spat back out) for the $\Delta < 0$ regime is different in Bohmian mechanics compared to ordinary quantum theory, and that their experimental results indicate a disagreement with the former. This claim is one part sleight of hand, and one part mistake.

First, they write that the dwell time is "identically defined and measurable in both Bohmian mechanics and standard quantum mechanics," but that it "yields different values in the two theories." Here is the sleight of hand: they say that the dwell time is identically defined because they use the equation $\tau_{\mathrm{dwell}} = N/j_{\mathrm{in}}$ in both cases (where $N$ is the number of stored particles, equal to integrating the squared amplitude over the high potential region), but they use a different definition of $j_{\mathrm{in}}$ for the two theories. Thus, what is being measured is really two different quantities.

For Bohmian mechanics, they define $j_{\mathrm{in}}$ proportional to the phase gradient, so that this measures the net probability current which goes into the guidance law for the Bohmian velocities. The stationary prediction for this value, of course, is zero, giving an infinite dwell time; just another way of stating that the Bohmian particles are frozen in the classically forbidden region. The experimenters measure the phase gradient and report it as a Bohmian velocity $v_{\mathrm{S}}$, so their Bohmian dwell time is:

$$\tau_{\mathrm{dwell,BM}} = \frac{N}{|\psi|^2 v_{\mathrm{S}}} \tag{13}$$

Note that the experimenters state their reported value of $v_{\mathrm{S}} = 59 \pm 42\ \mathrm{km} \cdot \mathrm{s}^{-1}$ as being "consistent with zero" and therefore with an infinite dwell time. We will examine this claim momentarily.

For standard quantum mechanics, they instead define $j_{\mathrm{in}}$ as proportional to the squared amplitude of the wavefunction times the velocity that the incident wave packet would have prior to the reflection process, which may be determined by analyzing the interference pattern before the potential step. Thus, they are using the expression:

$$\tau_{\mathrm{dwell,QM}} = \frac{4N}{|\psi_{\mathrm{max}}|^2 v_{\mathrm{in}}} \tag{14}$$

With $v_{\text{in}} = k_0/m$. This definition assumes that the net probability current (which is zero) is the sum of two equal and opposite contributions, attributable to the incident and reflected waves (that accounts for one factor of 2; the other factor of 2 comes from the fact that the probability density fluctuates in the region before the step, so the average probability density before the step is half the maximum). Of course, since Bohmian mechanics reproduces the particle position distribution of standard quantum theory, it gives the same prediction for the "dwell time" if it is defined using this version of $j_{\text{in}}$.

There are conceptual problems with this definition, over and above the bait-and-switch. The very concept of dwell time, and the split of the net probability current into equal and opposite contributions, supposes that the particle in some way travels towards the potential step, enters the classically forbidden region, dwells there for a while, and then turns around and leaves, all while the wavefunction is (quasi-)stationary. But ordinary quantum mechanics countenances no such thing: instead, it says there is nothing more to the particle than the wavefunction – and the wavefunction is not behaving like the imagined particle in the dwell time scenario at all. For most of the reflection process, the wavefunction in the vicinity of the potential step consists of a standing wave before the step-up and an exponentially decaying tail after the step-up, neither of which are moving.

Moreover, the split of the probability current into incident and reflected parts is mathematically justifiable only in the region before the step-up. Thus, it is arbitrary to suppose that this current flows into the high potential region, rather than (say) being immediately reflected at the step-up. A further indication of its arbitrariness is the fact that this definition only works in cases where the wavefunction has a well-defined wavelength, rather than a more general structure. So, the "dwell time" is not a valid physical concept in standard quantum mechanics.

Now, Sharoglazova et. al. raise the concern that an infinite dwell time would distort the observed distribution of the particles, since the Bohmian particles would enter the classically forbidden region, become trapped indefinitely, and then escape the optical cavity by transmitting through the mirror towards the camera. There is validity to this concern: the finite lifetime of the particles, roughly 0.27 ns, is much smaller than the duration of the experiment, 26 ns. So, the particles cannot have zero velocity for most of the duration of the experiment in the way that Bohmian mechanics (supposedly) predicts.

Their mistake at this point is taking one Bohmian prediction, made via the assumption of a stationary quantum state that neglected the finite particle lifetime, and considering it to be unchanged for the situation where the particle losses are an important factor. In fact, when we consider the particle losses, there is no problem for Bohmian mechanics.

We can model particle losses by adding an imaginary potential, proportional to the decay rate $\gamma$, to the Schrodinger equation as shown below. This makes the evolution of the wavefunction non-unitary; the total probability density for a single particle decreases over time as the probability that it decays accumulates.

$$
\begin{aligned}
i\frac{\partial \psi_m}{\partial t} &= -\frac{1}{2m}\frac{\partial^2 \psi_m}{\partial x^2} + V_m\psi_m + V_i\psi_a - \frac{i\gamma}{2}\psi_m \\
i\frac{\partial \psi_a}{\partial t} &= -\frac{1}{2m}\frac{\partial^2 \psi_a}{\partial x^2} + V_a\psi_a + V_i\psi_m - \frac{i\gamma}{2}\psi_a
\end{aligned} \tag{15}
$$

This changes the wavefunction dynamics. Simulations of the process with the loss rate included may be seen in Figures 3 and 4. Most relevantly for this discussion, in the $\Delta < 0$ regime, the reflected wave is more affected by the decay than the incident wave (as it has travelled a longer path from the source region). This makes the reflected wave lower in amplitude, and the result is a net current transporting probability amplitude along the waveguide. This is notably different than the zero current in the lossless case, even though the overall probability distribution is similar.

The guidance law for the Bohmian particles also changes; in addition to the guiding velocity along the waveguides and the jump rate between the waveguides, there is now a probability that the Bohmian particles will decay out of the system. If they decay with rate $\gamma$, the continuity equations for probability are now:

$$
\begin{aligned}
\frac{\partial \rho_m}{\partial t} + \frac{\partial}{\partial x}(\rho_m v_m) &= \sigma_a\rho_a - \sigma_m\rho_m - \gamma\rho_m \\
\frac{\partial \rho_a}{\partial t} + \frac{\partial}{\partial x}(\rho_a v_a) &= \sigma_m\rho_m - \sigma_a\rho_a - \gamma\rho_a
\end{aligned} \tag{16}
$$

And it is easy to check that $(\rho_m, \rho_a) = (|\psi_m|^2, |\psi_a|^2)$ is a solution to (16) if $(\psi_m, \psi_a)$ is a solution to (15). So, Bohmian mechanics still produces the same empirical predictions as ordinary quantum mechanics. In this scenario, we would expect the Bohmian dwell time to be roughly equal to the particle lifetime, since that is the rate at which they are being lost.

We can use the experimenters' measurement of the phase gradient to estimate the actual dwell time of the particles. Note from the simulation in Figure 4 that the probability density near $x = 0$, where they measure the phase gradient, is roughly the same size as $|\psi_{\mathrm{max}}|^2$. Then we may rearrange equations (13) and (14) to find:

$$
\tau_{\mathrm{dwell,BM}} \approx \frac{v_{\mathrm{in}}}{4v_{\mathrm{S}}}\tau_{\mathrm{dwell,QM}} \tag{17}
$$

Using values of $v_{\mathrm{S}} = 59 \pm 42\ \mathrm{km}\cdot\mathrm{s}^{-1}$, $v_{\mathrm{in}} \approx 4200\ \mathrm{km}\cdot\mathrm{s}^{-1}$, and $\tau_{\mathrm{dwell,QM}} \approx 2.5\ \mathrm{ps}$ (estimating $v_{\mathrm{in}} = \sqrt{2E/m}$ and $\tau_{\mathrm{dwell,QM}}$ from near the average $\Delta$ where $v_{\mathrm{S}}$ was measured), we get a very rough estimate for the Bohmian dwell time between 30 and 130 ps. This is close to the particle lifetime of 270 ps, at least in terms of orders of magnitude. Moreover, running a simulation of the reflection process with particle losses (see Figure 4), we get a Bohmian velocity of roughly $v_{\mathrm{S}} \approx 42\ \mathrm{km}\cdot\mathrm{s}^{-1}$ and a dwell time $\tau_{\mathrm{dwell,BM}} \approx 310\ \mathrm{ps}$. More precise analysis may be needed to be sure, but this is roughly consistent with both the experimental data and the expectation that the dwell time will be close to the particle lifetime.

**Conclusion**

It has been shown above that Bohmian mechanics necessarily makes the same predictions as ordinary quantum mechanics for the distribution of particle positions in the experiment presented in [1], even when the finite particle lifetime is considered. Far from being challenged by the experiment, Bohmian mechanics demonstrates that this indirect speed measurement is simply not valid, since the theory is able to predict the particle distribution used to infer the speed, without the particles having that speed.

Moreover, rather than confirming that the Bohmian particles have infinite dwell time in the classically forbidden region, which would be impossible given the finite lifetime of the photons in the optical cavity, the measurements of the phase gradient in [1] appear to be consistent with the Bohmian dwell time being roughly equal to the particle lifetime, which matches what is expected from the Bohmian perspective when the finite particle lifetime is accounted for in the analysis.

**Acknowledgement**

Thank you to Travis Norsen and Jan Klaers for helpful discussion.

**Bibliography**

[1] Sharoglazova, V., Puplauskis, M., Mattschas, C. *et al.* Energy–speed relationship of quantum particles challenges Bohmian mechanics. *Nature* **643**, 67–72 (2025). https://doi.org/10.1038/s41586-025-09099-4

[2] Norsen, T., Lande, J., McKagen, S. B. How and why to think about scattering in terms of wave packets instead of plane waves. (2009). https://doi.org/10.48550/arXiv.0808.3566

[3] Norsen, T. The pilot-wave perspective on quantum scattering and tunneling. *Am. J. Phys.* **81**, 258 (2013). https://doi.org/10.1119/1.4792375

[4] Durr, D., Goldstein, S., Tumulka, R., Zanghi, N. Bell-Type Quantum Field Theories. *J. Phys. A.* **38** R1 (2005). https://doi.org/10.1088/0305-4470/38/4/R01

[5] Norsen, T. On the explanation of Born-rule statistics in the de Broglie-Bohm pilot-wave theory. *Entropy* **20**(6), 422 (2018). https://doi.org/10.3390/e20060422

[6] Wang, Y., Wang. X., Wang, H., Lu, C. Tunnelling photons pose no challenge to Bohmian mechanics. (2025). https://doi.org/10.48550/arXiv.2507.20101

[7] Klaers, J., Sharoglazova, V., Toebes, C. Particle motion associated with wave-function density gradients. *Phys. Rev. A.* **107**, 052201 (2023). https://doi.org/10.1103/PhysRevA.107.052201

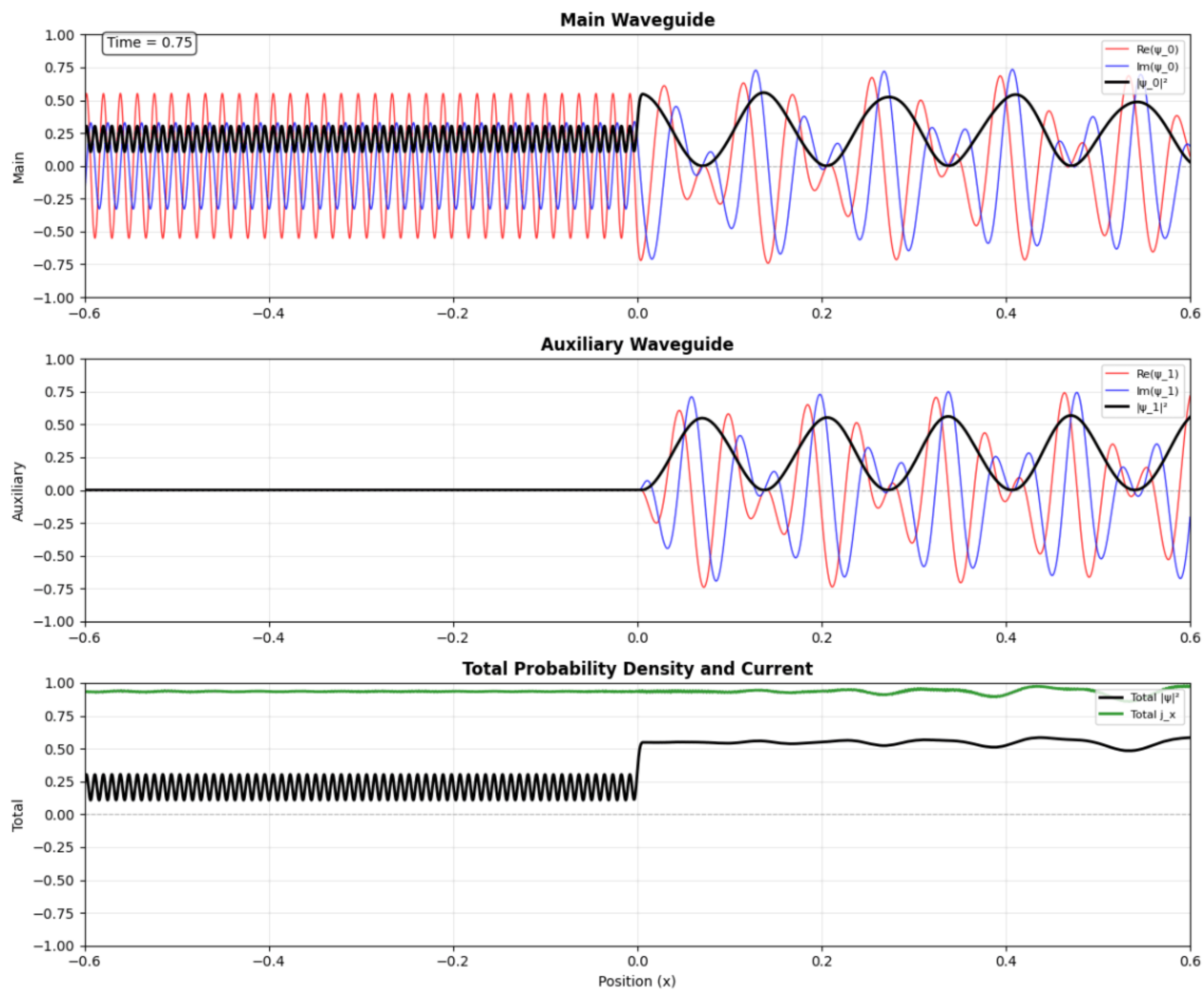

**Figure 1:** Simulation of an incident packet with $\Delta = 2.5J_0$ transmitting through the potential step, then tunnelling between the main and auxiliary waveguides. Particle losses are not modelled. (Units are mm for length and ns for time, with $\hbar = 1$.)

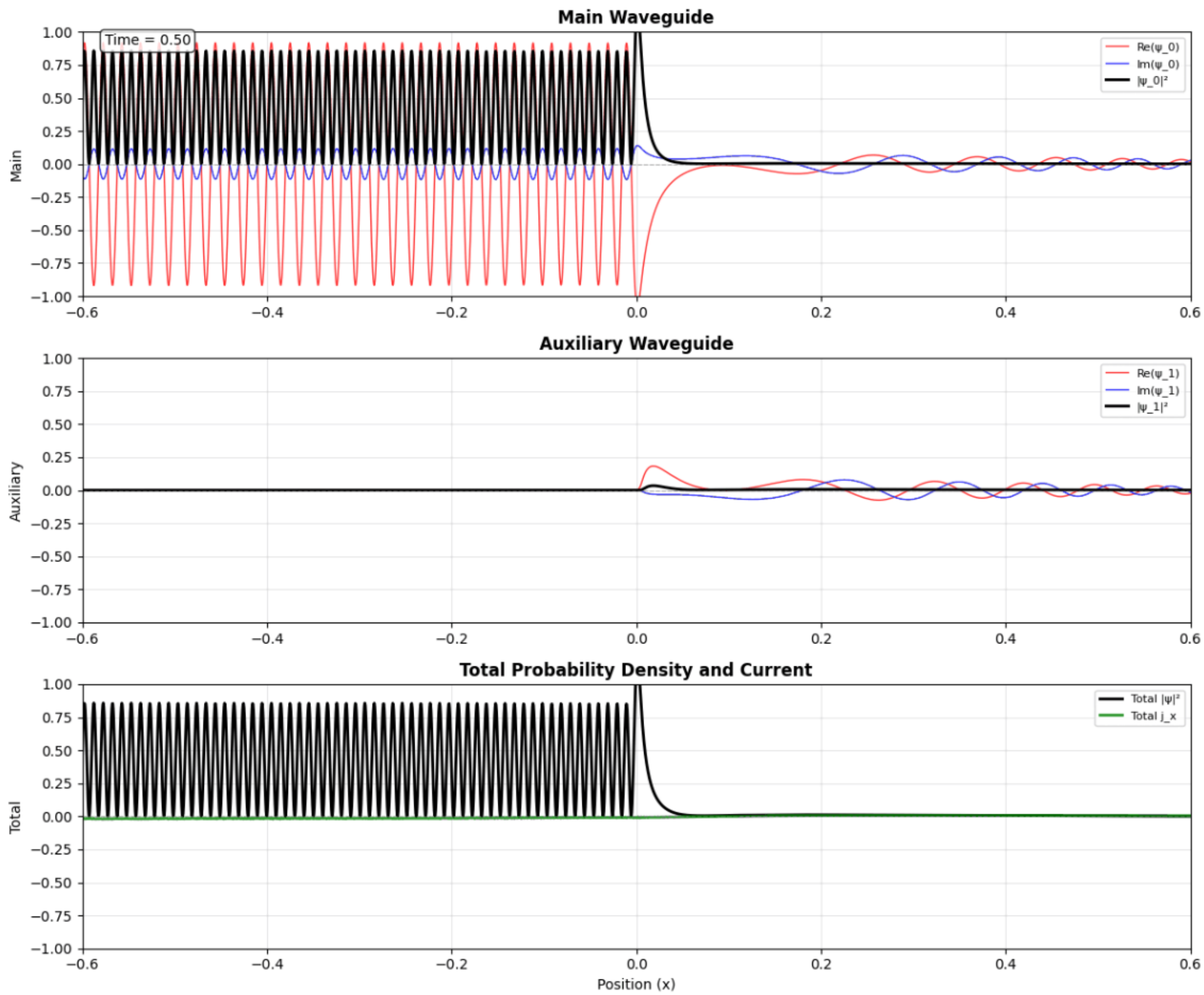

**Figure 2:** Simulation of an incident packet with $\Delta = -1.25J_0$ reflecting off the potential step almost completely, with only a small portion of the leading edge penetrating the high potential region. Particle losses are not modelled. Note the probability current is zero across the whole domain.

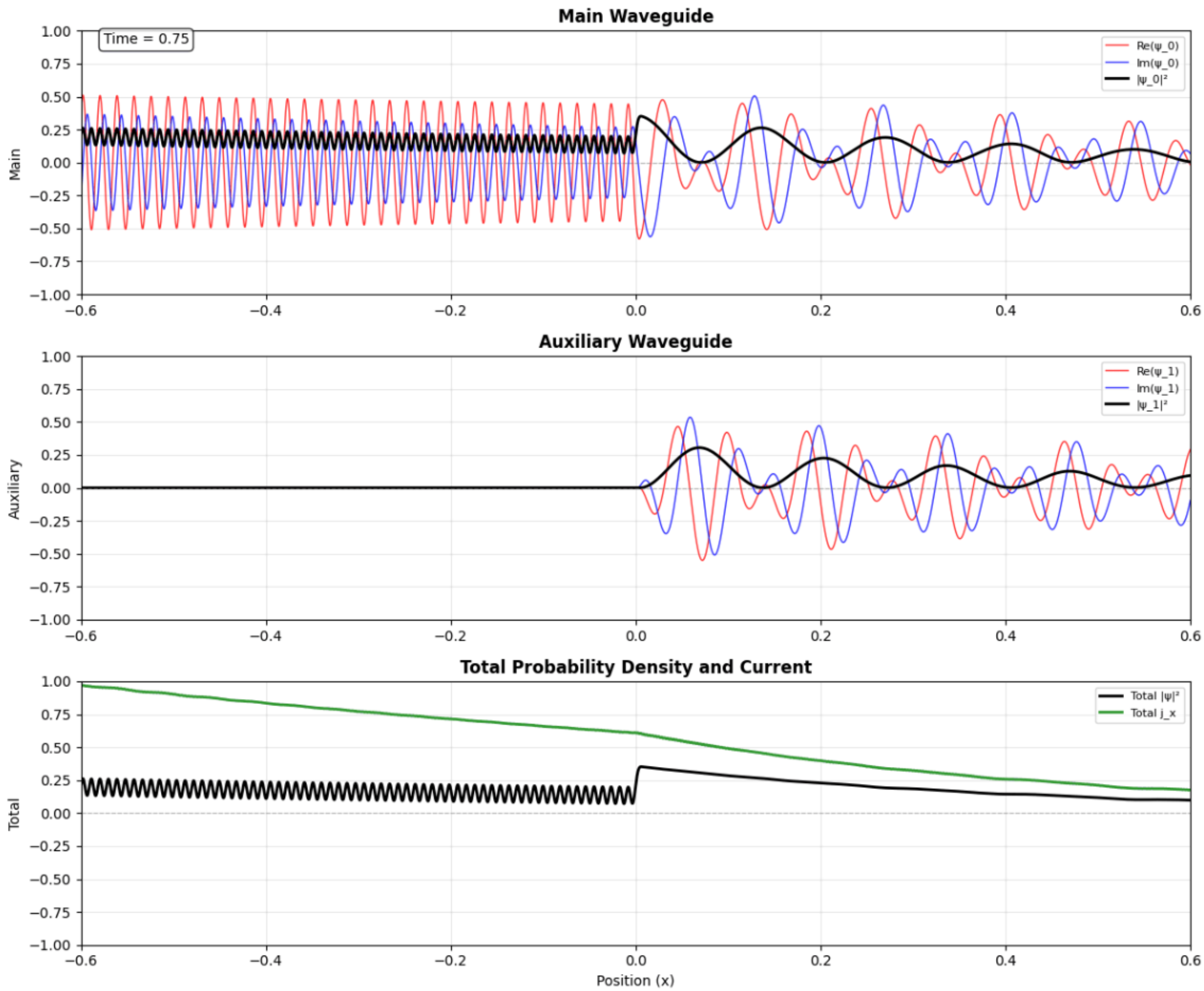

**Figure 3:** As in Figure 1, but with particle losses included in the simulation, using the decay rate $\gamma = 3.7\ \mathrm{ns}^{-1}$. The probability density and current diminish to the right as more of the probability amplitude is absorbed by the decay process.

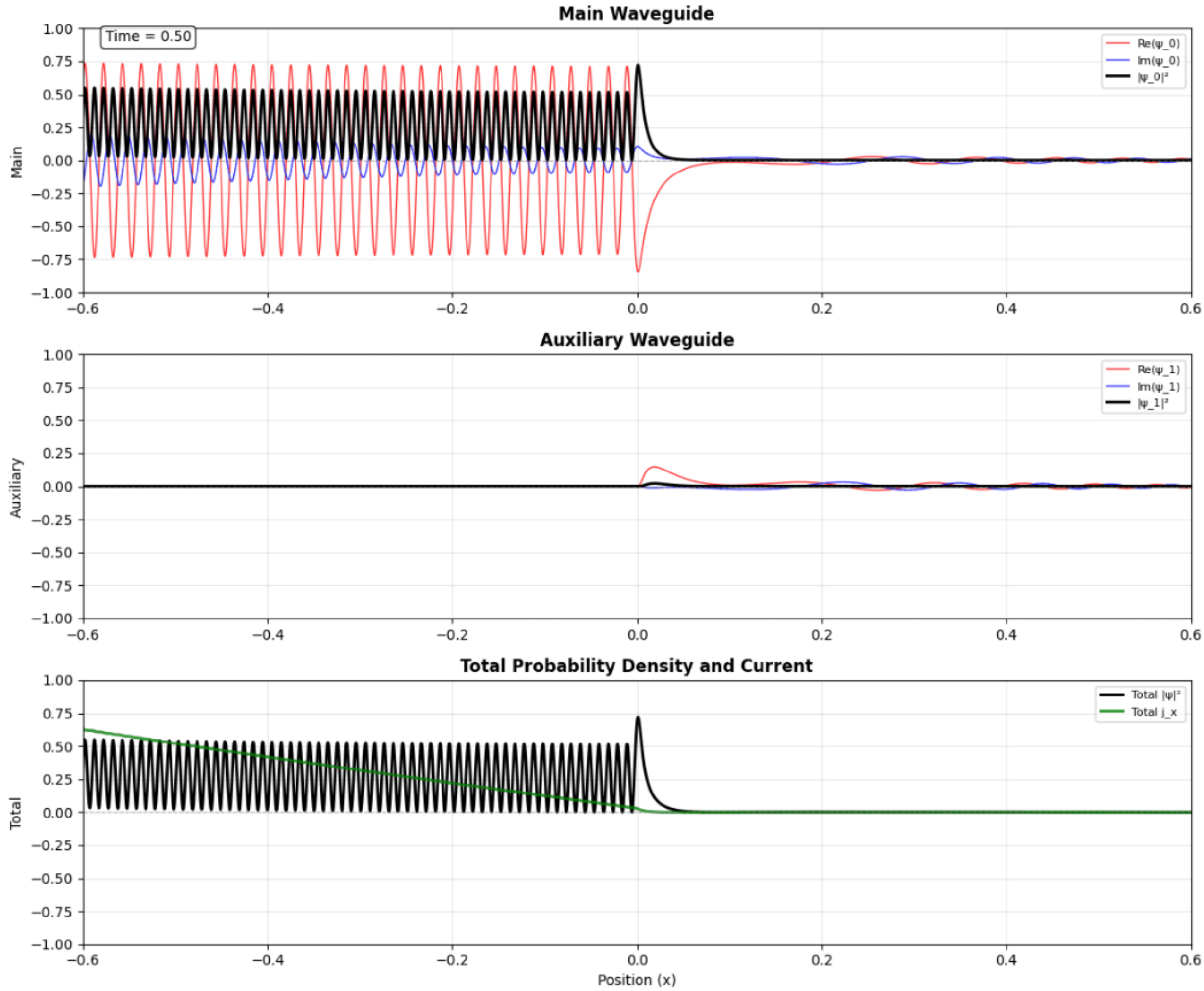

**Figure 4:** as in Figure 2, but with particle losses included in the simulation. Note the net current transporting probability amplitude along the waveguide to replace what is being absorbed by the decay process, in contrast to the zero current in the lossless case.

At $x = 0$, this simulation has a probability current of $0.029\ \mathrm{ns}^{-1}$ and a probability density of $0.69\ \mathrm{mm}^{-1}$, while the integral of the probability density over the $x > 0$ region is $N = 0.009$. This implies $v_\mathrm{S} \approx 42\ \mathrm{km \cdot s^{-1}}$ and $\tau_{\mathrm{dwell,BM}} \approx 310\ \mathrm{ps}$.

**Appendix A**

We may solve the time-independent Schrodinger equation (6) for the system:

$$E\psi_m = -\frac{1}{2m}\frac{\partial^2\psi_m}{\partial x^2} + V_m\psi_m + V_i\psi_a$$

$$E\psi_a = -\frac{1}{2m}\frac{\partial^2\psi_a}{\partial x^2} + V_a\psi_a + V_i\psi_m$$

For the $x < 0$ region, we have $\psi_a(x) = 0$ and $\psi_m(x)$ behaves as a free particle:

$$E\psi_m = -\frac{1}{2m}\frac{\partial^2\psi_m}{\partial x^2}$$

Which has the following solution (here we always have $E > 0$):

$$\psi_m(x) = C_I e^{ik_0 x} + C_R e^{-ik_0 x}, \qquad k_0 = \sqrt{2mE}$$

For the $x > 0$ region, note (with credit to the authors of [6]) that:

$$E(\psi_m + \psi_a) = -\frac{1}{2m}\frac{\partial^2}{\partial x^2}(\psi_m + \psi_a) + V_0(\psi_m + \psi_a)$$

$$E(\psi_m - \psi_a) = -\frac{1}{2m}\frac{\partial^2}{\partial x^2}(\psi_m - \psi_a) + (V_0 - 2J_0)(\psi_m - \psi_a)$$

This is easily solved (supplying a factor of 2 on the coefficients for convenience):

$$\psi_m(x) + \psi_a(x) = 2A_+ e^{ik_+ x} + 2B_+ e^{-ik_+ x}, \qquad k_+^2 = 2m(E - V_0)$$

$$\psi_m(x) - \psi_a(x) = 2A_- e^{ik_- x} + 2B_- e^{-ik_- x}, \qquad k_-^2 = 2m(E - V_0 + 2J_0)$$

Which can be rearranged to:

$$\psi_m(x) = A_+ e^{ik_+ x} + B_+ e^{-ik_+ x} + A_- e^{ik_- x} + B_- e^{-ik_- x}$$

$$\psi_a(x) = A_+ e^{ik_+ x} + B_+ e^{-ik_+ x} - A_- e^{ik_- x} - B_- e^{-ik_- x}$$

Note that we may also write $k_\pm^2 = 2m(\Delta \mp J_0)$ with $\Delta = E - V_0 + J_0$. This solution has three different regimes. If $\Delta > J_0$, both $k_+$ and $k_-$ are real (and we may take them to be positive), and we have two propagating modes. If $\Delta < -J_0$, both $k_+$ and $k_-$ are imaginary, and we have two exponential modes. In that case we may write $k_\pm = i\kappa_\pm$ and take $\kappa_\pm$ to be positive reals. Finally, if $|\Delta| < J_0$, we have one exponential mode $k_+ = i\kappa_+$ and one propagating mode $k_-$.

Considering $\Delta > J_0$, we assume no leftwards propagating modes and require $\psi_a(0) = 0$ for continuity with the $x < 0$ region, which eliminates the $B_\pm$ coefficients and sets $A_\pm$ equal:

$$\psi_m(x) = A\left(e^{ik_+x} + e^{ik_-x}\right) = 2A\cos(k_1x)\,e^{ik_2x}$$

$$\psi_a(x) = A\left(e^{ik_+x} - e^{ik_-x}\right) = -2iA\sin(k_1x)\,e^{ik_2x}$$

$$k_+ = \sqrt{2m(\Delta - J_0)}\,, \qquad k_- = \sqrt{2m(\Delta + J_0)}$$

$$k_1 = \frac{k_- - k_+}{2}\,, \qquad k_2 = \frac{k_+ + k_-}{2}$$

Continuity of $\psi_m$ and $\partial\psi_m/\partial x$ at $x = 0$ determine the following relationships:

$$C_I = \left(1 + \frac{k_+ + k_-}{2k_0}\right)A = \left(1 + \frac{k_2}{k_0}\right)A$$

$$C_R = \left(1 - \frac{k_+ + k_-}{2k_0}\right)A = \left(1 - \frac{k_2}{k_0}\right)A$$

Here we see the reflected wave has lower amplitude than the incident wave. We may calculate the early spatial structure of $p_a(x)$:

$$p_a(x) = \frac{|\psi_a(x)|^2}{|\psi_m(x)|^2 + |\psi_a(x)|^2} = \sin^2(k_1x) = k_1^2x^2 + O(x^4)$$

For $\Delta < -J_0$, the requirement that the wavefunction must exponentially decay towards the right (and not exponentially blow up) eliminates the $B_\pm$ coefficients again, and continuity with $\psi_a(0)$ again sets $A_\pm$ equal:

$$\psi_m(x) = A(e^{-\kappa_+x} + e^{-\kappa_-x}) = 2A\cosh(\kappa_1x)\,e^{-\kappa_2x}$$

$$\psi_a(x) = A(e^{-\kappa_+x} - e^{-\kappa_-x}) = -2A\sinh(\kappa_1x)\,e^{-\kappa_2x}$$

$$\kappa_+ = \sqrt{2m(|\Delta| + J_0)}\,, \qquad \kappa_- = \sqrt{2m(|\Delta| - J_0)}$$

$$\kappa_1 = \frac{\kappa_+ - \kappa_-}{2}\,, \qquad \kappa_2 = \frac{\kappa_+ + \kappa_-}{2}$$

Then from continuity of $\psi_m$ and $\partial\psi_m/\partial x$ at $x = 0$:

$$C_I = \left(1 + i\frac{\kappa_+ + \kappa_-}{2k_0}\right)A = \left(1 + i\frac{\kappa_2}{k_0}\right)A$$

$$C_R = \left(1 - i\frac{\kappa_+ + \kappa_-}{2k_0}\right)A = \left(1 - i\frac{\kappa_2}{k_0}\right)A$$

Showing that the magnitudes of the reflected and incident waves are the same, so the reflection is total. Calculating $p_a(x)$, we find:

$$p_a(x) = \operatorname{sech}(2\kappa_1x)\sinh^2(\kappa_1x) = \kappa_1^2x^2 + O(x^4)$$

For $|\Delta| < J_0$, the same requirements as before result in the solution:

$$\psi_m(x) = A\left(e^{-\kappa_+ x} + e^{ik_- x}\right)$$

$$\psi_a(x) = A\left(e^{-\kappa_+ x} - e^{ik_- x}\right)$$

$$\kappa_+ = \sqrt{2m(J_0 - \Delta)}\,, \qquad k_- = \sqrt{2m(J_0 + \Delta)}$$

And the relationship between the coefficients:

$$C_I = \left(1 + \frac{k_- + i\kappa_+}{2k_0}\right)A\,, \qquad C_R = \left(1 - \frac{k_- + i\kappa_+}{2k_0}\right)A$$

With this we can find the spatial structure of $p_a(x)$ for the intermediate regime:

$$p_a(x) = \frac{1}{2}\left(1 - \operatorname{sech}(\kappa_+ x)\cos(k_- x)\right) = \frac{1}{4}(\kappa_+^2 + k_-^2)x^2 + O(x^4)$$

Note that the cases $|\Delta| = J_0$ can be treated as degenerate versions of the $|\Delta| < J_0$ case, with the exponentially decaying $\kappa_+$ mode becoming constant if $\Delta = J_0$, and the propagating $k_-$ mode becoming constant if $\Delta = -J_0$.

We can put the results for the small $x$ spatial structure of $p_a$ from the three regimes together to infer the apparent speed, using equation (3):

$$v = \begin{cases} \dfrac{J_0}{\sqrt{m\left(|\Delta| - \sqrt{|\Delta|^2 - J_0^2}\right)}}, & |\Delta| > J_0 \\ \sqrt{J_0/m}\,, & |\Delta| \le J_0 \end{cases}$$

When $|\Delta/J_0| \gg 1$ we have an approximate relationship:

$$|\Delta| - \sqrt{|\Delta|^2 - J_0^2} \approx \frac{J_0^2}{2|\Delta|}$$

(in the sense that the ratio of these quantities rapidly approaches 1 as $|\Delta|$ increases), so that the apparent speed satisfies:

$$v \approx \sqrt{2|\Delta|/m}$$

Which is the theoretical prediction stated by the authors of [1].

**Appendix B**

We can show the equivariance property of Bohmian mechanics (that $|\psi|^2$- distributed particles remain $|\psi|^2$-distributed) for this system by substituting $(\rho_m, \rho_a) = (|\psi_m|^2, |\psi_a|^2)$ into the evolution equations (12), and verifying that they hold:

$$\frac{\partial |\psi_m|^2}{\partial t} + \frac{\partial}{\partial x}(|\psi_m|^2 v_m) =^? \sigma_a |\psi_a|^2 - \sigma_m |\psi_m|^2$$

$$\frac{\partial |\psi_a|^2}{\partial t} + \frac{\partial}{\partial x}(|\psi_a|^2 v_a) =^? \sigma_m |\psi_m|^2 - \sigma_a |\psi_a|^2$$

Working things out for $\psi_m$ (everything follows similarly for $\psi_a$), the time derivative of $|\psi_m|^2$ comes from the Schrodinger equation (4):

$$\frac{\partial |\psi_m|^2}{\partial t} = \psi_m^* \frac{\partial \psi_m}{\partial t} + \psi_m \frac{\partial \psi_m^*}{\partial t} = \frac{i}{2m}\psi_m^* \frac{\partial^2 \psi_m}{\partial x^2} - \frac{i}{2m}\psi_m \frac{\partial^2 \psi_m^*}{\partial x^2} - iV_i \psi_m^* \psi_a + iV_i \psi_a^* \psi_m$$

Multiplying $|\psi_m|^2$ by $v_m$ (10) gives the probability current, and the spatial derivative of this expression cancels the second derivative terms:

$$\frac{\partial}{\partial x}(|\psi_m|^2 v_m) = -\frac{i}{2m}\frac{\partial}{\partial x}\left(\psi_m^* \frac{\partial \psi_m}{\partial x} - \frac{\partial \psi_m^*}{\partial x}\psi_m\right) = -\frac{i}{2m}\psi_m^* \frac{\partial^2 \psi_m}{\partial x^2} + \frac{i}{2m}\psi_m \frac{\partial^2 \psi_m^*}{\partial x^2}$$

Leaving us with the following terms in the equation we want to verify:

$$-iV_i \psi_m^* \psi_a + iV_i \psi_a^* \psi_m =^? \sigma_a |\psi_a|^2 - \sigma_m |\psi_m|^2$$

Substituting in the expressions for $\sigma_a$ and $\sigma_m$ (11):

$$\sigma_a |\psi_a|^2 - \sigma_m |\psi_m|^2 = [iV_i(\psi_a^* \psi_m - \psi_m^* \psi_a)]^+ - [iV_i(\psi_m^* \psi_a - \psi_a^* \psi_m)]^+$$

Now since $[f]^+ = f$ for $f > 0$ and otherwise $[f]^+ = 0$, if we have real-valued $f$, then:

$$[f]^+ - [-f]^+ = f$$

So:

$$[iV_i(\psi_a^* \psi_m - \psi_m^* \psi_a)]^+ - [iV_i(\psi_m^* \psi_a - \psi_a^* \psi_m)]^+ = iV_i(\psi_a^* \psi_m - \psi_m^* \psi_a)$$

Which is just what we needed to verify the evolution equation.

Note that when we add the loss rate term to the Schrodinger equation (15), we end up with an additional term of $-\gamma |\psi_m|^2$ in the expression for $\partial |\psi_m|^2 / \partial t$ (similarly for $\psi_a$), which is exactly balanced if we add a decay rate of $\gamma$ for the Bohmian particles (probability $\gamma dt$ for the particle to decay in any time interval $dt$) as in equation (16).